\documentclass[aps,pra,twocolumn,amsmath,amssymb,showpacs]{revtex4}
\usepackage{graphicx}% Include figure files
\usepackage{epsfig}
\usepackage{dcolumn}% Align table columns on decimal point
\usepackage{bm}% bold math

\usepackage{amssymb}
\usepackage{bm}
\usepackage{amsmath}

\begin{document}

\title{Vacuum Rabi oscillation induced by virtual photons in the ultrastrong coupling
regime}
\author{C. K. Law}
\affiliation{Department of Physics and Institute of Theoretical
Physics, The Chinese University of Hong Kong, Shatin, Hong Kong
Special Administrative Region, People's Republic of China}

\begin{abstract}
We present an interaction scheme that exhibits a dynamical
consequence of virtual photons carried by a vacuum-field dressed
two-level atom in the ultrastrong coupling regime. We show that,
with the aid of an external driving field, virtual photons provide
a transition matrix element that enables the atom to evolve
coherently and reversibly to an auxiliary level accompanied by the
emission of a real photon. The process corresponds to a type of
vacuum Rabi oscillation, and we show that the effective vacuum
Rabi frequency is proportional to the amplitude of a single
virtual photon in the ground state. Therefore the interaction
scheme could serve as a probe of ground state structures in the
ultrastrong coupling regime.

\end{abstract}

%42.50.-p, 03.65.Nk, 42.65.-k
%42.50.-p Quantum optics
%03.65.Nk Scattering theory
%42.65.-k Nonlinear optics
\maketitle

A single-mode electromagnetic field interacting with a two-level
atom has been a fundamental model in quantum optics capturing the
physics of resonant light-matter interaction. In particular, the
Jaynes-Cummings (JC) model \cite{JC1,JC2}, which describes the
regime where the interaction energy $\hbar \lambda$ is much
smaller than the energy scale of an atom $\hbar \omega_A$ and a
photon $\hbar \omega_c$, has tremendous applications in cavity QED
\cite{Kimble_review,Haroche} and trapped ion systems
\cite{Wineland}. Recently, there has been considerable research
interest in the ultrastrong coupling regime where $\lambda$
becomes comparable to $\omega_c$ and $\omega_A$. Such a regime has
been explored by experiments in various related systems with
artificial atoms and cavity photon resonators, including
superconducting qubit in coplanar waveguide \cite{Niemczyk} or LC
resonator \cite{BSshift}, microcavities embedding doped quantum
wells \cite{Gunter,Todorov}, and two-dimensional electron gas
coupled to metamaterial resonators \cite{Scalari}. In addition,
theoretical investigations have also found novel phenomena in the
ultrastrong coupling regime, such as the asymmetry of vacuum
Rabi-splitting \cite{Cao}, photon blockade \cite{Ridolfo},
nonclassical states generation \cite{Ashhab1}, superradiance
transition \cite{Ashhab2}, and collapse and revivals dynamics
\cite{Casanova}.

A key feature in the ultrastrong coupling regime is the
significant number of virtual photons existing around the
vacuum-field dressed atom. These virtual photons are generated by
counter-rotating terms in the Hamiltonian, and they can have
direct physical consequences. For example, by modulating the
atom-field coupling strength virtual photons can be released as a
form of quantum vacuum radiation \cite{Liberato}. In this paper we
address a different effect of the vacuum-field dressed atom,
namely, a kind of vacuum Rabi oscillations that would not occur if
virtual photons are absent.

Specifically, we investigate the quantum dynamics of a driven
quantum Rabi model. The configuration of our system is shown in
Fig. 1 in which a $\Xi$-type  three-level atom is confined in a
single-mode cavity. The atomic levels $|g \rangle $ and $| e
\rangle$ are coupled to a cavity field of frequency $\omega_c$.
These two atomic levels and the cavity field mode constitute a
Rabi model. In addition, there is an external classical field
driving the transition between $|e \rangle$ and the third atomic
level $|f \rangle$. We note that some theoretical aspects of
three-level artificial atoms in circuit QED was discussed in
\cite{YouNori}, and $\Xi-$type superconducting atoms have been
demonstrated in experiments \cite{3level_1,3level_2,3level_3}.
Recently Carusotto {\it et al.} have studied the dynamics of a
related system in a different driving configuration
\cite{Carusotto}.

The Hamiltonian of our system is given by $(\hbar =1)$,
\begin{equation}
H = H_{R} + \omega_f |f \rangle \langle f|+\Omega \cos \omega_p t
\left({  |f \rangle \langle e| +  |e \rangle \langle f|} \right)
\end{equation}
where $H_R$ is the Hamiltonian of the Rabi model \cite{Rabi},
\begin{equation} H_R = \frac
{\omega_0} {2} (|e \rangle \langle e| - |g \rangle \langle g|) +
\omega_c a^\dag a + \lambda (a+a^\dag) (|g \rangle \langle e|+
|e\rangle \langle g|).
\end{equation}
Here $\omega_0$ is the (bare) transition frequency between $|e
\rangle $ and $| g \rangle$, and $\omega_f-\omega_0/2$ is the
transition frequency between $|f \rangle$ and $| e \rangle$. The
parameter $\lambda$ denotes the atom-cavity coupling strength, and
the classical driving field has a frequency $\omega_p$ and an
interaction strength $\Omega$. In writing $H_R$, we have kept
counter-rotating terms because $\lambda$ is comparable to
$\omega_c$ in the ultrastrong coupling regime. Note that the
coupling between the cavity mode and the level $|f \rangle$ is
assumed to be weak and so that it is not included in the
Hamiltonian.

\begin{figure}[ptb]
\center
\includegraphics[width=2.2 in]{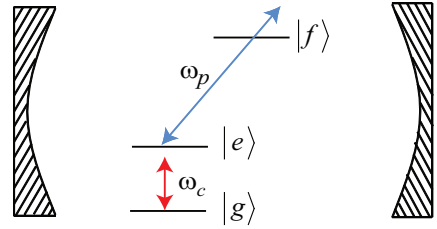}
\caption{(Color online) Interaction scheme of a $\Xi-$type
three-level atom in a cavity. The atomic states $|g \rangle$ and
$|e \rangle$ and a cavity field mode of frequency $\omega_c$ form
a quantum Rabi model described by $H_R$, and an external classical
field of frequency $\omega_p$ drives the transition between $|e
\rangle$ and $|f \rangle$.}\label{setup}
\end{figure}

Initially the system is prepared in the ground state of $H_R$,
which is the lowest-energy state of the system in the absence of
the driving field. Our task is to determine the dynamics after the
driving field is turned on. To analyze the problem, we apply a
unitary transformation to simplify the Hamiltonian. It is known
that for low energy states of the Rabi model, $H_R$ can be
transformed to into a form of Jaynes-Cummings Hamiltonian
approximately by a unitary operator $e^{-S}$ \cite{Gan}. Here the
operator $S$ and its parameters are defined by:
\begin{eqnarray}
&& S=\frac{ \lambda \xi} {\omega_c} (|g \rangle \langle e|+
|e\rangle \langle g|) (a^\dag-a),
\\
&&\xi = \frac {\omega_c} {\omega_c+ \eta \omega_0},  \\
&& \eta =  \exp(-\frac{2\lambda^2\xi^2}{\omega_c^2}).
\end{eqnarray}
Then it can be shown that $H_R' =  e^S H_R e^{-S}$ is
approximately given by \cite{Gan,Zheng1,Zheng2,Oh}
\begin{eqnarray}
H_R' & \approx & \frac {\omega_0'} {2} (|e \rangle \langle e| - |g
\rangle \langle g|) +  \omega_c a^\dag a + \lambda' (a |e\rangle
\langle g|+a^\dag |g \rangle \langle e|)
\nonumber \\
&& +\frac{\lambda^2 \xi } {\omega_c} (\xi-2) (|e \rangle \langle
e| + |g \rangle \langle g|) \nonumber \\
& \equiv &  H_{JC}
\end{eqnarray}
where $H_{JC}$ describes a JC model in which the atomic frequency
and cavity-atom interaction strength are renormalized as
$\omega_0' = \eta \omega_0$ and $\lambda' = {2 \eta \omega_0 \xi
\lambda} / {\omega_c} $, respectively.

Note that $H_{JC}$ in Eq. (6) is an approximation to $H_R'$, and
the difference $H_R' - H_{JC}$ describes multi-photon processes
that correspond to higher order corrections
\cite{Gan,Zheng1,Zheng2,Oh}. Since $|g,0 \rangle$ is the ground
state of $H_{JC}$, $e^{-S} |g,0 \rangle$ is an approximated ground
state of $H_R$ in the original frame. The accuracy of such an
approximation has been tested in Ref. \cite{Gan}. Specifically, if
$\lambda$ is comparable but smaller than $\omega_c$, the ground
state energy of $H_{JC}$ has a good agreement with that of $H_R$
obtained by exact numerical calculations over a range of
parameters. For example in the case $\omega_c=\omega_0= 2
\lambda$, the approximated ground state energy obtained by
$H_{JC}$ has the percentage error about $0.65\%$.

Now we perform the transformation for our system Hamiltonian $H$,
which becomes,
\begin{eqnarray}
H' &=& e^S H e^{-S} \nonumber \\
& \approx & H_{JC} + \omega_f |f \rangle \langle f| \nonumber
\\ & & + \Omega \cos \omega_p t ( e^{S} |e \rangle \langle f|+ |f \rangle \langle e|e^{-S}).
\end{eqnarray}
Since $e^S |e \rangle = \cosh [\frac {\lambda \xi } {\omega_c}
(a^\dag-a)] |e \rangle + \sinh [\frac {\lambda \xi } {\omega_c}
(a^\dag-a)] |g \rangle$, we expand the hyperbolic sine and cosine
operator functions in normal order up to first order in $\lambda
\xi / \omega_0$,
\begin{eqnarray}
\cosh \left[ {\frac{{\lambda \xi }}{{\omega _c }}(a^\dag   - a)}
\right] &\approx& \eta^{1/4},
\\
\sinh \left[ {\frac{{\lambda \xi }}{{\omega _c }}(a^\dag   - a)}
\right] &\approx & \eta^{1/4} {\frac{\lambda \xi} {\omega_c} (
a^{\dag} - a} )
\end{eqnarray}
Therefore the transformed Hamiltonian becomes,
\begin{eqnarray}
H'  & \approx &  H_{JC} + \omega_f |f \rangle \langle f| + \Omega'
\cos \omega_p t
 \left( { |f\rangle \langle e|
+|e\rangle \langle f|}\right) \nonumber
\\
&&  +  \frac{\lambda \xi}{\omega_c} \Omega' \cos {\omega_p t}
\left({  |g \rangle \langle f| -  |f \rangle \langle g| } \right)
( a^\dag- a )
\end{eqnarray}
where $\Omega' = \eta^{1/4} \Omega$ is a renormalized driving
field strength, and the last term indicates a new coupling between
$|g \rangle $ and $| f \rangle$ through the cavity field mode.

A further simplification can be made by exploiting resonance when
$\omega_p$ is tuned to a certain resonance frequency defined by
the undriven system. In this paper we consider the resonance at
\begin{equation}
\omega_p = \omega_f +\omega_c  -\left[ \frac{\lambda^2 \xi }
{\omega_c} (\xi-2)- \frac {\omega_0'} {2} \right],
\end{equation}
which corresponds to the transition between $|g,0 \rangle$ to $|f,
1\rangle$, since the square bracket term is the approximate ground
state energy of $H_R$ by the transformation method. By the
condition (11), $|g,0 \rangle$ and $|f, 1\rangle$ are resonantly
coupled, but $|f, 1\rangle$ and $|e, 1\rangle$ is far away from
resonance (the corresponding detuning is of order $\omega_c$).
Therefore if $\Omega'$ is not too strong, the system is confined
to the two resonantly coupled states, i.e., all off-resonant
transitions may be ignored. In this way $H'$ in the interaction
picture is reduced to
\begin{eqnarray}
H'_I  & \approx &   -\frac{\lambda \xi}{2 \omega_c} \Omega'
\left({ |g,0 \rangle \langle f,1| +  |f,1 \rangle \langle g,0| }
\right).
\end{eqnarray}
Eq. (12) indicates that the system would execute a form of vacuum
Rabi oscillations, in which $|g,0 \rangle$ behaves as an excited
atom in the vacuum field, and $|f,1 \rangle$ behaves as an ground
atom with a single photon. In cavity QED, such oscillations lead
to vacuum Rabi splitting \cite{VR1,VR2,VR3}. Note that the
effective vacuum Rabi frequency here is ${\lambda \xi} \Omega' /{
\omega_c}$, which is significant in the ultrastrong coupling
regime where $\lambda$ is comparable to $\omega_c$.

It is useful to go back to the original frame in which the Rabi
oscillations occur between the states $e^{-S} |f,1 \rangle$ and
$e^{-S} |g,0 \rangle$. Since $e^{-S} |f,1 \rangle = |f,1 \rangle$,
an initial ground state will evolve to $|f,1 \rangle$ after half
of a Rabi period. If we switch off the external field at this
moment, the single photon described by $|f,1 \rangle$ will be free
to escape the cavity because the atom in the state $|f \rangle$
does not couple to the cavity field when $\Omega=0$, i.e., the
photon cannot be reabsorbed by the atom. In this way,  a $\pi$
pulse of the driving field can generate a real photon
deterministically while the atom is excited to the $|f \rangle$
state.

To gain a better insight of the physical process without relying
on the approximation made in Eqs. (6) and (10), we express the
Hamiltonian by the eigenbasis of $H_R$. Let $|\psi_n \rangle$ be
an eigenvector of $H_R$ with the eigenvalue $\lambda_n$, i.e.,
$H_R | \psi_n \rangle = \lambda_n  | \psi_n \rangle$ (the ground
state is denoted by $|\psi_0 \rangle$), and consider the expansion
$|e, n \rangle = \sum_m c_{nm} | \psi_m \rangle$ with the
coefficients $c_{nm} = \langle \psi_m |e, n \rangle$. Therefore
\begin{equation}
|f \rangle \langle e| = \sum_n |f,n \rangle \langle e,n| =
\sum_{nm} c_{nm}^* |f,n \rangle  \langle \psi_m|. \\
\end{equation} In this way, the Hamiltonian (1) in the interaction
picture becomes,
\begin{equation}
H_I = \Omega  \cos \omega_p t \sum_{nm} e^{i (\omega_f+ n\omega_c
- \lambda_m) t} c_{nm}^* |f,n \rangle \langle \psi_m| + h.c.
\end{equation}
At the resonant frequency $\omega_p = \omega_f +\omega_c -
\lambda_0$, $|\psi_0 \rangle$ and $|f,1 \rangle$ are resonantly
coupled. If we keep only the resonant terms, then we have
\begin{equation} H_I \approx \frac {\Omega c_{10}^*} {2}
|f,1 \rangle \langle \psi_0| + h.c.
\end{equation}
Comparing with $H_I'$ in Eq. (12) and noting that $| \psi_0
\rangle \approx e^{-S} |g,0 \rangle$, $H_I$ describes the same
type of resonant interaction as $H_I'$. However, we emphasize that
$H_I$ in Eq. (15) is a more accurate interaction Hamiltonian than
$H_I'$ because $H_I$ is derived directly from the eigenbasis of
$H_R$ without making use of the approximation in Eq. (6). In this
sense, the resonant condition (11) can be improved by replacing
the square bracket term by $\lambda_0$.

\begin{figure}[ptb]
\center
\includegraphics[bb=20 1 332 189, width=2.8 in]{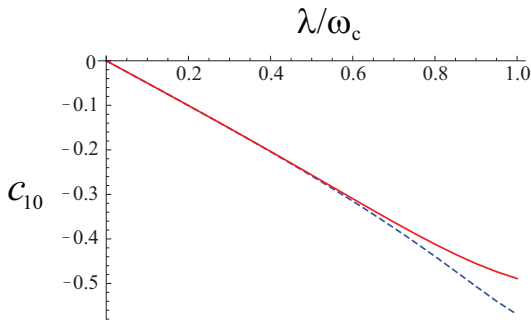}
\caption{(Color online) Probability amplitude of $|e,1 \rangle$ in
the ground state of $H_R$ as a function of the coupling strength
$\lambda$ for the $\omega_0=\omega_c$ case. The solid red line
corresponds to exact numerical values, and the dashed blue line is
obtained from the approximated ground state $e^{-S}|g,0 \rangle$
according to Eq. (6). }
\end{figure}

The role of virtual photons is now explicitly seen in Eq. (15)
through the effective vacuum Rabi frequency ${\Omega |c_{10}}|$.
This is because $c_{10}$ is precisely the probability amplitude of
a single virtual photon state in $|\psi_0 \rangle$. In other
words, we may interpret that the interaction described in Eq. (15)
is induced or mediated by a virtual photon. In Fig. 2, we plot
$c_{10}$ (solid line) as a function of $\lambda/ \omega_c$ for the
case $\omega_c= \omega_0$, and the figure shows that the magnitude
of $c_{10}$ is appreciable in the ultrastrong coupling regime. As
a comparison, we also plot the approximate amplitude $c_{10}
\approx -\eta^{1/4} \xi \lambda/\omega_c$  (dashed line) obtained
from $e^{-S} |g,0 \rangle$. For the parameters used in Fig. 2, we
see that the approximation agrees well with the exact numerical
calculation up to $\lambda/ \omega_c < 0.6$.

We have tested our prediction of the virtual-photon-induced Rabi
oscillations by solving numerically the Schr\"odinger equation
defined by the Hamiltonian (1) with the initial state $|\psi_0
\rangle$. In Fig. 3 we plot the exact numerical probability
$P_{1f}$ of the system in the state $|f, 1 \rangle$ as a function
of time. The parameter $\lambda = \omega_c/2$ used in the figure
is served as an example of ultrastrong coupling. We see the Rabi
cycles as predicted by the Hamiltonians (12) or (15) for
relatively weak driving fields with $\Omega \le 0.4 \omega_c$. At
a stronger driving field with $\Omega = 0.8 \omega_c$ (red solid
line), and there is a high frequency pattern due to counter
rotating terms of the classical driving field, and the Rabi
oscillations are less perfect in the sense that the maximum
$P_{1f} \approx 0.9$ is smaller than one. Such a behavior is
understood because the off-resonance transitions neglected in Eq.
(12) or (15) would generate energy shifts which in turn could
bring the driven system out of resonance. As a result, the
amplitude of oscillations in $P_{1f}$ is reduced. Since these
energy shifts are generally proportional to $\Omega^2$,  as long
as $\Omega$ is small compared with detunings associated with
off-resonance transitions, it would be safe to use Eq. (15), and
this is demonstrated in Fig. 3 for $\Omega$ up to $0.4 \omega_c$.

\begin{figure}[ptb]
\center
\includegraphics[bb=20 1 332 230, width=2.8 in]{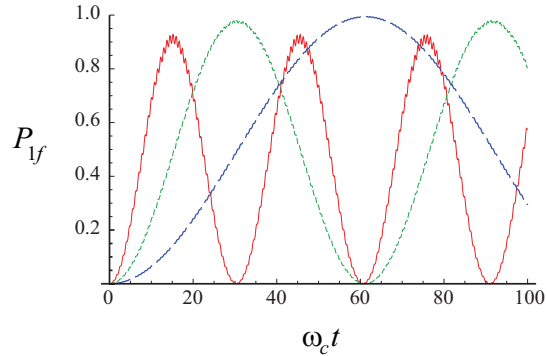}
\caption{(Color online) Probability of $|f,1 \rangle$ as a
function of time for $\Omega=0.2 \omega_c$(blue long dashed), $0.4
\omega_c$(green short dashed) and  $\ 0.8 \omega_c$ (red solid).
The parameters used are: $\lambda=0.5 \omega_c$,
$\omega_c=\omega_0= \omega_f/3$, $\omega_p = \omega_f +\omega_c -
\lambda_0$, and the numerical ground state energy $\lambda_0 =
-0.633 \omega_c$. The figure is essentially the same if $\omega_p$
in Eq. (11) is used.}
\end{figure}
Finally, it is worth noting that the Hamiltonian in Eq. (14) has
higher resonances at $ \omega_p = \omega_f+ n\omega_c - \lambda_0$
for odd positive integers $n$. The requirement of an odd $n$ is
because $|\psi_0 \rangle$ has a definite parity in which the
atomic state $|e \rangle$ and odd photon numbers are connected. In
the case $n=3$, the driving field at the corresponding $\omega_p$
would resonantly excite the atom to $|f \rangle$ with the emission
of three real photons. The effective Hamiltonian would be of the
same form of (15), but with $|f,1\rangle$ and $c_{10}^*$ replaced
by $|f,3\rangle$ and $c_{30}^*$, i.e., the effective Rabi
frequency is proportional to $|c_{30}|$. Such a three-photon
resonance was also observed in our numerical calculations.

To conclude, we have shown that virtual photons in the ultrastrong
coupling regime can play a key role in quantum dynamics by
providing the transition matrix elements that allow the system to
access relevant quantum states of interest. In our scheme, the
system can exhibit a form of vacuum Rabi oscillations which can be
considered as a signature of virtual photons. Since our main focus
in this paper is on the interaction induced by virtual photons,
decoherence effects have not been included in the discussion.
However, as long as the decoherence times is sufficiently short,
coherent dynamics predicted by the Hamiltonian (12) or (15) would
be justified. Specifically, given a vacuum Rabi period $T \approx
2 \pi \omega_c /\lambda \xi \Omega'$, the cavity field damping
rate $\gamma_c$ and atomic decay rate $\gamma_A$, the condition
$\gamma_j T \ll 1$ ($j=c, A$) ensures that the system can execute
a Rabi cycle without being affected by the damping, and this is
achievable in the ultrastrong coupling regime with moderate small
$\gamma$'s. For the parameters used in Fig. 3, for example,
$\gamma_j < 10^{-2} \omega_c$ would be sufficient. We emphasize
that a finite interaction time within $T$ is of practical
importance, since the interaction (12) or (15) is switchable via
the driving field. This feature could be a tool for performing
quantum operations on qubits formed by the atom or the field, as
well as for deterministic single-photon generation
\cite{Law,Kimble,Grangier}. In addition, since the effective
vacuum Rabi frequency is proportional to the corresponding virtual
photon amplitude, our scheme can be used to probe the ground state
structure of the quantum Rabi model.

\begin{acknowledgments}
The author thanks Dr. H. T. Ng for discussions. This work is
partially supported by a grant from the Research Grants Council of
Hong Kong, Special Administrative Region of China (Project
No.~CUHK401812).
\end{acknowledgments}

 \narrowtext

\end{document}